%% file: no-grammar.tex
\documentclass[preprint, journal]{vgtc}                % final (journal style)
\ifpdf%                                % if we use pdflatex
  \pdfoutput=1\relax                   % create PDFs from pdfLaTeX
  \usepackage{graphicx}                % allow us to embed graphics files
  \DeclareGraphicsExtensions{.pdf,.png,.jpg,.jpeg} % for pdflatex we expect .pdf, .png, or .jpg files
\else%                                 % else we use pure latex
  \usepackage{graphicx}                % allow us to embed graphics files
  \DeclareGraphicsExtensions{.eps}     % for pure latex we expect eps files
\fi%

\graphicspath{{figures/}{pictures/}{images/}} % where to search for the images

\usepackage{microtype}                 % use micro-typography (slightly more compact, better to read)
\PassOptionsToPackage{warn}{textcomp}  % to address font issues with \textrightarrow
\usepackage{textcomp}                  % use better special symbols
\usepackage{mathptmx}                  % use matching math font
\usepackage{times}                     % we use Times as the main font
\usepackage{cite}                      % needed to automatically sort the references
\usepackage{tabu}                      % only used for the table example
\usepackage{booktabs}                  % only used for the table example
\usepackage{amssymb}
\usepackage{adjustbox}
\usepackage{array}
\usepackage{ stmaryrd }
\usepackage{multibib}
\usepackage{wrapfig}
\usepackage{fontawesome}
\newcites{appendix}{Other Sources}
\onlineid{0}

\vgtccategory{Research}
\vgtcpapertype{theory/model}

\usepackage{xcolor}
\usepackage{hyperref}
\input{commands}
\input{system-commands.tex}

\title{No Grammar to Rule Them All: \\ A Survey of JSON-style DSLs for Visualization}

\author{Andrew M. McNutt}
\authorfooter{
 Andrew M. McNutt is with University of Chicago. E-mail: mcnutt@uchicago.edu.
}

\shortauthortitle{McNutt: No Grammar}

\abstract{
There has been substantial growth in the use of JSON-based grammars, as well as other standard data serialization languages, to create visualizations. 
Each of these grammars serves a purpose: some focus on particular computational tasks (such as animation), some are concerned with certain chart types (such as maps), and some target specific data domains (such as ML). 
Despite the prominence of this interface form, there has been little detailed analysis of the characteristics of these languages. 
In this study, we survey and analyze the design and implementation of \numLangs{} JSON-style DSLs for visualization.
We analyze these languages supported by a collected corpus of examples for each DSL (consisting of \numExamples{} instances) across a variety of axes organized into concerns related to domain, conceptual model, language relationships, affordances, and general practicalities.
We identify tensions throughout these areas, such as between formal and colloquial specifications, among types of users, and within the composition of languages. 
Through this work, we seek to support language implementers by elucidating the choices, opportunities, and tradeoffs in visualization DSL design. 
}


\CCScatlist{ % not used in journal version
 \CCScat{K.6.1}{Management of Computing and Information Systems}%
{Project and People Management}{Life Cycle};
 \CCScat{K.7.m}{The Computing Profession}{Miscellaneous}{Ethics}
}

\vgtcinsertpkg

\begin{document}


\firstsection{Introduction}

\maketitle

Domain-specific languages (DSLs) represented in standard data serialization formats, such as JSON or YAML, are an increasingly common\cite{pu2021special} interface for the specification of visualizations across an array of contexts and tasks.
These restricted textual languages allow for the declarative specification of both static and interactive graphics in a systematic manner that can be manipulated both by humans, making them attractive for end-user programming, and computational agents, making them appealing for artificial intelligence applications\cite{wu2021ai4vis}.
This language style appears in a surprisingly large variety of tools and systems but is well exemplified by Vega\cite{satyanarayan2014declarative} and Vega-Lite \cite{satyanarayan2016vega}.

While it is sometimes derided for usability issues\cite{KoHatesJson, fowler2010domain,liuatlas}, this language style has a variety of benefits. DSLs which employ it can be \emph{expressive}, allowing for the concise manipulation of complex specifications with minimal textual modification\cite{satyanarayan2014declarative}.
Many of these languages enhance the \emph{explorability} of a space of possible programs by simple and fluid movement between instances.
Their limited scope enables some specifications to be used \emph{portably}, such that charts created in one platform (such as a GUI like Voyager~\cite{wongsuphasawat_voyager_2017}) can be used in another environment
(such as in the Python-based Altair~\cite{vanderplas_altair_2018}).

Despite the popularity (\figref{fig:over-time}) of this approach, there has been little detailed analysis of the characteristics of these systems.
Pu \etal{} \cite{pu2021special} highlight the need for additional study of visualization grammars,
while Wongsuphasawat\cite{wongsuphasawat2020encodable} surveyed the more general space of  JavaScript (JS) visualization libraries.
Although they are insightful, these works leave critical questions about these DSLs open:
\emph{What problems do they seek to solve? Who are they designed to serve?} or more generally \emph{What design and implementation patterns are used in JSON-style DSLs?}

In this paper we answer these questions by surveying visualization DSLs represented in standard data serialization languages covering academic, industrial, and open source language efforts, yielding \numLangs{} distinct languages (\figref{fig:family-table}).
We analyze each of these DSLs across a variety of dimensions including the motivations for their design, relationships with other languages, and the conceptual models which are utilized.
We identify five sets of concerns (\figref{fig:analysis-structure}) which are critical to JSON-style DSL design and highlight a corresponding set of tensions to be navigated,
such as the tension between formal and colloquial models or the effect on the DSL caused by the interplay of different intended users.
In doing so we note opportunities, tradeoffs, and open challenges.
To aid this analysis we collected examples of each DSL, yielding  \numExamples{} programs, available in our \asLink{\liveurltext}{interactive supplement}.

While JSON-style DSLs have been usefully employed in a variety of visual analytics systems\cite{satyanarayan_critical_2020, wongsuphasawat_voyager_2017,zong2021lyra,mcnutt2021integrated},
we believe that a firmer grasp of the design space of this language form will help future languages better address the highlighted design questions.
Moreover, a language's affordances, abstractions, and models guide the types of expression that are made using it\cite{orchard2011four,MediaForThinkingTheUnthinkable}.
Thus, a stronger foundation may open the door to new forms of analysis and expression.

\section{Related Work}\label{sec:related-work}

We will now locate our study within prior work on DSLs in general (and review relevant terminology) and visualization DSLs specifically.

\subsection{Domain Specific Languages}

DSLs are a type of programming language designed to facilitate particular tasks within a chosen domain.
While this term is variably defined, DSLs are usually (but not always\cite{fowler2010domain}) defined as languages that are unable to execute general computations, in exchange for specific declarative notation related to a domain of interest---properties that differentiate them from General Purpose Languages (GPLs).
The database query language SQL, the browser-styling language CSS, and the markup language \LaTeX{} are all familiar examples of this design approach.
% https://stackoverflow.com/questions/2968411/ive-heard-that-latex-is-turing-complete-are-there-any-programs-written-in-late
Van Deursen \etal{} \cite{van2000domain} argue that DSLs are useful because they allow domain experts to operate within the notation of a given domain, and assert that they are typically concise, reusable, and self-documenting.
They also note that DSLs carry a host of disadvantages including maintenance costs, learnability issues, and the danger of \emph{language cacophony}\cite{fowler2010domain} resulting from a preponderance of languages.

DSLs are often thought of as solely declarative, as the user specifies intent relative to the domain rather than through low-level details of how that action is executed.
However, some DSLs use imperative syntax (\eg{} shaders or Atlas\cite{liuatlas}), but all DSLs in our study are declarative and the closest exceptions are pipeline models.

A common decomposition of DSLs\cite{mernik2005and, fowler2010domain} describes them as \emph{external} or \emph{internal}.
\emph{External} languages define their syntax outside of their host language, such that they typically require separate parsing to execute.
Some utilize custom syntax (\eg{} SQL or CSS), while others elect to use standard data serialization grammars (\eg{} XML or JSON).

This raises questions about what qualifies as a language as opposed to an API.
Highlighting this ambiguity, Fowler\cite{fowler2010domain} argues for a heuristic related to a fluent, composable, or language-like nature.
This refers to the concept that ``expressiveness comes not just from individual expressions, but also from the way they can be composed together''\cite{fowler2010domain}.
That is, there is a systemic way that the language operates, without needing special cases for every expression form.
We use this heuristic to identify languages in our survey.

The complement to \emph{external} DSLs are \emph{internal} DSLs,  which are embedded (as a library or through syntax extensions) into a host language---\eg{} dplyr, d3, or RSpec.
These languages provide expressivity similar to external languages, but do so in a way that confers the benefits and limitations of their host.
Tobin-Hochstadt \etal{}\cite{tobin2011languages} highlight the permeable border between languages and libraries in Racket, where libraries are distributed as language extensions---an ambiguity that is increasingly relevant as more visualization libraries adopt language-style interfaces.
This architectural choice allows for the straightforward creation of richly expressive languages that are easy to integrate into a host, but can force constraints and notation which are inappropriate to the DSL domain.
In contrast, Diderot\cite{kindlmann2015diderot} explicitly resists embedding so as to maintain the domain specificity of its type system.

Prior studies have sought to understand and typify DSL usage\cite{fowler2010domain, van2000domain, poltronieri2021usability,borum2021designing}.
Mernik \etal{}~\cite{mernik2005and} describe design patterns exhibited at each stage of the DSL design process.
Van Deursen \etal{}\cite{van2000domain} characterize 75 DSLs by purpose.
The analysis of our survey draws on these works but is designed to complement these considerations made by others by focusing on a particular domain.
Erdweg \etal{}\cite{erdweg2012language} identify a set of language composition mechanisms, which guides our discussion of the topic (although ours is adapted to a less-general domain).
Several studies consider DSLs in specific domains, such as declarative data analytics \cite{makrynioti2019declarative}, configuration languages \cite{gunther2012software}, and visual computing\cite{shen2021domain}.
Our work is related to these but is centered on visualization.

\begin{figure}[t]
  \centering
  \includegraphics[width=\linewidth]{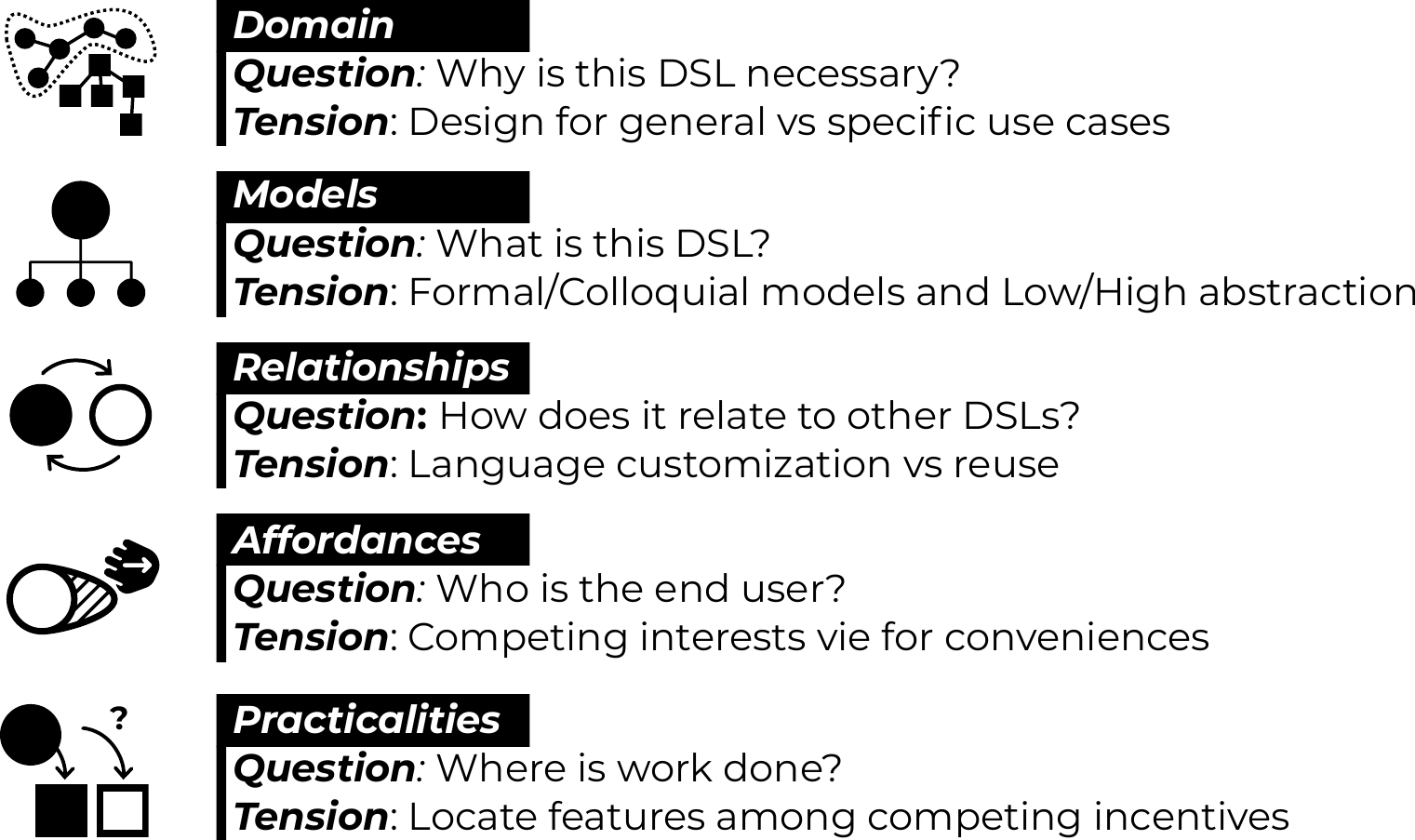}
  \caption{
    An overview of the analysis of our survey.
  }
  \label{fig:analysis-structure}
  \vspace{-0.2in}
\end{figure}

\begin{figure}[t]
  \centering
  \includegraphics[width=0.9\linewidth]{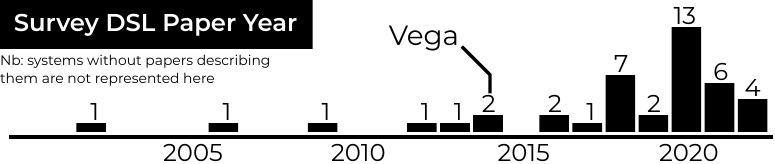}
  \caption{
    Since \vega{}'s publication JSON-style DSLs have become popular.
  }
  \label{fig:over-time}
  \vspace{-0.2in}
\end{figure}

The use of standard serialization languages as carrier languages for DSLs is not new.
XML and other hierarchical serialization languages have long been used as a way to configure applications\cite{fowler2010domain, gunther2012software} and even specify visualizations (as in VizML\cite{wilkinson2012grammar}).
Similarly, visualization is not unique in its use of JSON-DSLs.
Beyond familiar uses such as configuration or NoSQL languages (\eg{} MongoDB), they are used in domains as varied as statistical analysis\cite{jun2019tea}, web development\cite{Varv22Borowski}, narrative\cite{compton2015tracery} and game generation\cite{duplantis2021genre}, chatbots\cite{klopfenstein2018adapting}, dance\cite{payne2021danceon}, and fabrication\cite{tran2021grammar}.
% \am{Microfluidic biochips?????} %https://mediatum.ub.tum.de/doc/1550266/file.pdf
We are interested in building a better understanding of JSON-style DSLs precisely because they are so prevalent---although their sudden prominence (\figref{fig:over-time}) may indicate that they are a fad.

% also cue, Jsonnet,
% Projects such as dhall\cite{Dhall21Design} try to alleviate some of these issues by providing variables and simple functions as a way to make refactoring of configuration files less error prone (although the advantages such a structure has over a more familiar GPL appear limited).
% https://www.lucidchart.com/techblog/2018/07/16/why-json-isnt-a-good-configuration-language/
% https://dhall-lang.org/#
% https://jsonnet.org/articles/comparisons.html

Despite their popularity, JSON-based languages are no panacea.
They are sometimes maligned for their lack of programming usability features\cite{KoHatesJson}, rendering them hard to learn, debug, and extend \cite{liuatlas}.
These criticisms can be extended through the Cognitive Dimensions of Notations (CDN)\cite{blackwell2003notational}, which are a suite of lightweight heuristics (\cogdim{highlighted} throughout) that characterize the usability of notational interfaces. For instance, \cogdim{viscosity} refers to the effort required to alter a program to a desired state, while \cogdim{diffuseness} describes how terse the language is.
This style of evaluation is especially useful for programming languages as it provides an external reference from which to critically reflect and a common grammar for usability issues.
Thus, we can add to the criticisms of JSON DSLs by noting that they are subject to errors related to \cogdim{premature commitment} (choices that make movement between states difficult), \cogdim{hard mental operations} (the work required outside of the coding environment), and \cogdim{progressive evaluation} (how incomplete programs are examined).
Some serialization formats (\eg{} YAML or JSON5) or languages (\eg{} dhall\cite{Dhall21Design}) seek to address usability issues---such as \cogdim{diffuse} syntax and lack of a \cogdim{secondary notation}. Yet, a consensus replacement has not materialized.
This may be due to JSON's ubiquity in modern systems, which impart rich error handling and parsing, as well as typings via projects like JSON Schema\cite{pezoa_foundations_2016}.

\subsection{Visualization DSLs}

Visualization features a rich space in which a DSL can usefully abstract away unnecessary details in favor of domain-appropriate notation.

The most prominent visualization DSL is Wilkinson's Grammar of Graphics (GoG) \cite{wilkinson2012grammar},
which describes the visualization process as a series of stages that results in a mapping of data attributes to visual-encoding channels (\eg{} a penguin's flipper length mapped to spatial position).
This approach allows the construction of myriad chart forms, in contrast to ``chart templates'' which map data attributes to aspects of a given chart type.
GoG has influenced the development of many contemporary visualization language systems \cite{wickham2010layered, bostock2011d3, satyanarayan2014declarative}.
Friendly\cite{Friendly22Colorless} reviews the model and its history.

Despite its prevalent use, the term  ``visualization grammar'' is not well defined and is used in a variety of ways\cite{pu2021special}.
This term is sometimes used in the generative syntactic sense\cite{park2017atom}, referring to a system of rules that can be repeatedly applied to create particular shades of meaning.
It may also be used to refer to composable systems of expression \cite{li2018p4, mcnutt2021integrated}, akin to Fowler's language definition.
Further, it may specifically refer to variants of Wilkinson's\cite{wilkinson2012grammar} GoG.
Still others use it to refer to any visualization system\cite{VizGrammar}.
We do not strive to provide a conclusive definition of visualization grammars here, instead electing to use the slightly more general framing of DSLs---although we sometimes use the term to refer to a space of allowed syntax.

There are a wide variety of DSLs for visualization that fall outside of our language form of interest.
The dot graph language and the mermaid diagramming language feature custom syntax for graph-based tasks.
APT\cite{mackinlay1986automating} is a DSL used to describe charts in a manner amenable to automated recommendations.
Idyll\cite{conlen2018idyll} is a DSL for visualization-mediated explorable explanations.
ViSlang\cite{rautek2014vislang} provides a system for making and coordinating small DSLs in SciVis, while
Diderot\cite{kindlmann2015diderot} uses notation specifically aligned with the tensor-calculus operations which arise in that setting.
Although the design patterns these DSLs manifest are valuable, they are beyond the scope of our study.

Several prior works study visualization languages.
Wongsuphasawat\cite{wongsuphasawat2020encodable, WongsuphasawatNavigating2020} sketched a taxonomy based on the level of abstraction covering graphics languages, low-level languages, grammars, high-level languages, and templating systems.
Qin \etal{} \cite{qin2020making} sketch a similar taxonomy based on an expressiveness-accessibility axis.
We expand upon these studies through a more in-depth survey of narrower scope.
Pu \etal{} \cite{pu2021special} highlight the pressing need for more formal study of these entities.
We seek to explore and address the questions they raise, as well as support future work by developing a richer understanding of the state of the art of this language form.
Satyanarayan \etal{} \cite{satyanarayan_critical_2020} reflect on the design of visualization authoring systems, the results of which overlap with our study, although tuned to a slightly different domain.

\begin{figure*}[t]
  \centering
  \includegraphics[width=\linewidth]{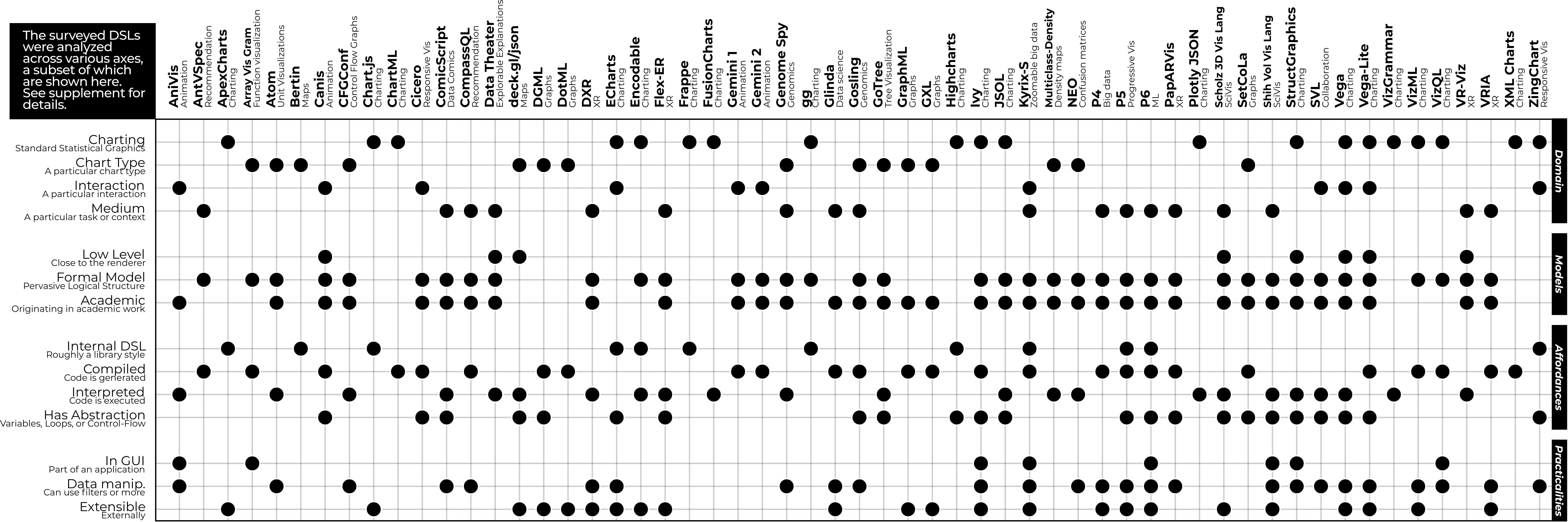}
  \caption{
    DSLs select different feature combinations to achieve their goals. No one language, or feature combination, will suit all situations.
  }
  \label{fig:family-table}
  \vspace{-0.2in}
\end{figure*}

\section{Survey Methodology}\label{sec:methodology}

\newcommand{\searchTerm}[1]{\texttt{#1}}

We conducted a survey of visualization languages represented fully or partially in standard serialization languages (\eg{} JSON, YAML, XML). This yielded \numLangs{} languages, which are displayed in \figref{fig:family-table}.

We searched relevant academic search engines (Google Scholar, ACM Digital Library, IEEE Xplore) and code repositories (GitHub) for the following \textbf{keywords}:
\searchTerm{DSL},  \searchTerm{domain-specific language}, \searchTerm{JSON}, \searchTerm{XML}, \searchTerm{YAML}, \searchTerm{visualization}, \searchTerm{map},  \searchTerm{grammar}, \searchTerm{language}, \searchTerm{chart}, and \searchTerm{graph}.
Given the influence of the works on \vega{} and \vegalite{} on this type of DSL,
we also reviewed all papers citing the papers documenting those systems \cite{satyanarayan2014declarative, satyanarayan_reactive_2016, satyanarayan2016vega}.
We utilized snowball sampling whenever possible.
We refer to systems in our survey like \sys{Vega}, while we cite the works documenting them.
See the appendix for a survey bibliography.

Our survey \textbf{criterion} included \emph{any human-usable language that uses a standard serialization language to produce visualizations.}
We follow Fowler's definition of a language\cite{fowler2010domain} as being a system with a concept of composition or a sense of fluency.
This \emph{language nature} can manifest in a variety of ways, such as mark or series composition, as well as data or view algebras.
We follow our prior definition\cite{mcnutt2021potential} of a visualization as being a transformation of data meant to be interpreted by a human.

This criterion excludes some system types.
SVG, HTML, and other high-level markup languages (as well as general-purpose JSON-based DSLs, such as Varv\cite{Varv22Borowski}) were excluded because they are capable of producing far more than visualizations.
Also excluded were those merely subsetting another language---for instance, GraphScape\cite{kim_graphscape_2017} uses a non-interactive subset of \vegalite{} to explore sequence recommendation.
These systems are excluded because they simply make use of a visualization DSL rather than constructing one.
Systems that possess systematically described languages (\eg{} Visception\cite{kristiansen2020visception}) but either do not utilize a standard serialization language for its description or do not expose that language to the end-user were excluded.
We focus on computer-based languages, which precludes natural language specifications such as in NL4DV\cite{narechania2020nl4dv}.
While many libraries excluded by our criterion (\eg{} ggplot) could be recast into JSON, we exclude them because  we strive only to understand the patterns of those DSLs that have explicitly opted to use this representation---although the design of visualization APIs more generally is intriguing future work.

\parahead{Examples.}
To facilitate comparison between the DSLs in our survey we collected representative samples of each language.
For some DSLs this involved collecting every single example available (such as \atom{} and \cicero{}). For those with thriving communities (such as \highcharts{} and \vegalite{}), we only gathered examples from documentation or test repositories which provided sufficient examples for analysis.
In some cases (\eg{} \gotree{}) the only examples available were those found in the publications documenting those languages.
While additional examples would always be useful, the samples collected were sufficiently representative to allow us to consider each of the axes of analysis.
This yielded \numExamples{} examples, although a small number of DSLs dominate this total.
We present these materials as a DSL zoo\cite{zaytsev2015grammar} in our supplement at \liveurl{} and for download at \osf{}.

\parahead{Analysis Process.}
We conducted an analysis seeking to answer: \emph{What are the design and implementation patterns in visualization DSLs represented in standard serialization formats?}
To do so, we analyzed each surveyed language across a set of topics.
Our initial selection of topics  was motivated by discussions of DSLs in general\cite{mernik2005and, van2000domain,fowler2010domain,liuatlas}, visualization DSLs \cite{wongsuphasawat2020encodable,liuatlas}, as well as related work on visualization authoring systems \cite{satyanarayan_critical_2020}.
We iteratively added and removed axes of analysis (analogous to codes) until a theoretical saturation was reached.
Each axis was evaluated based on available documentation (such as a paper describing the DSL), the found examples for the language, or sometimes by reviewing the code itself.
These results were grouped into categories (\figref{fig:analysis-structure}).
See the \asLink{\liveurltext}{supplement} for details.

Our observations and analysis are descriptive, and not evaluative. Thus analyses such as locating DSLs within Satyanarayan \etals{} \cite{satyanarayan_critical_2020} expressiveness-learnability spectrum are beyond the scope of this work, as such  comparisons would require experimental evaluation.
We focus on patterns relevant to visualization DSLs and refer to exterior sources for those related to general DSL patterns\cite{fowler2010domain, karsai2014design}.
We forgo analysis of living sources (such as interviews) because these artifacts are sufficiently rich to conduct our analysis, although future work could be augmented by such explorations.

\section{Analysis}

We organize our discussion following our five concerns (\figref{fig:analysis-structure}) guided by the information-gathering interrogatives for each DSL.

\subsection{Domain: Why is it necessary?}\label{sec:domain}

We begin by considering the aims of our DSLs in order to identify the problems they seek to solve and therein identify why they are necessary.

We found four purposes or domains for designing visualization DSLs:
creating standard charts, creating a particular chart, enabling a specific interaction, and serving a certain task--- as in \figref{fig:family-table}.
These various purposes highlight a critical tension: \emph{why not just use something that already exists?}
Indeed, many of the graphic types and domains can be addressed using  \vega{} or GoG\cite{Friendly22Colorless} with enough manipulation.
This issue can be seen as a \emph{Turing tar pit}\cite{perlis1982special} in which everything is possible, but nothing is easy.
The value of using DSLs is exactly to avoid this pitfall: allowing some things to be easy by making some things (that are not relevant to the domain) impossible.

\parahead{Standard charts.} Most DSLs focused on standard charts, such as bar charts or scatter plots.
Among these, the ostensible purpose varied, featuring different levels of abstraction, contexts, means of expression, or implementation affordances---each of which we discuss in subsequent sections.
For instance, \vegalite{} enables standard charting tasks using a formal GoG-inspired approach, while \echarts{} employs a colloquial chart-type model that uses its close connection to the browser to provide responsive and progressive analytics features\cite{li2018echarts}.
The \emph{purpose} of such languages is then the additional affordances brought to the design.

\parahead{Chart Forms.} Some DSLs focus on enabling a particular chart form or genre.
For instance, several languages support maps (such as \bertin{} and \deckgljson{}), while others focus on graphs (\graphml{}, \gxl{}, \cfgConf{}, and \setcola{}).
\neo{} enables confusion matrices.
By focusing on a specific chart form a DSL can tailor its notation to the concerns of that graphic.
To wit, graphs specify node position as relationships between entities and not in terms of spatial attributes. This allows languages like \graphml{} to focus on only high-level attributes---although adding notions of axial direction can be useful (which \setcola{} achieves by encoding directional and relative properties as constraints).

\begin{figure}[t]
  \centering
  \includegraphics[width=\linewidth]{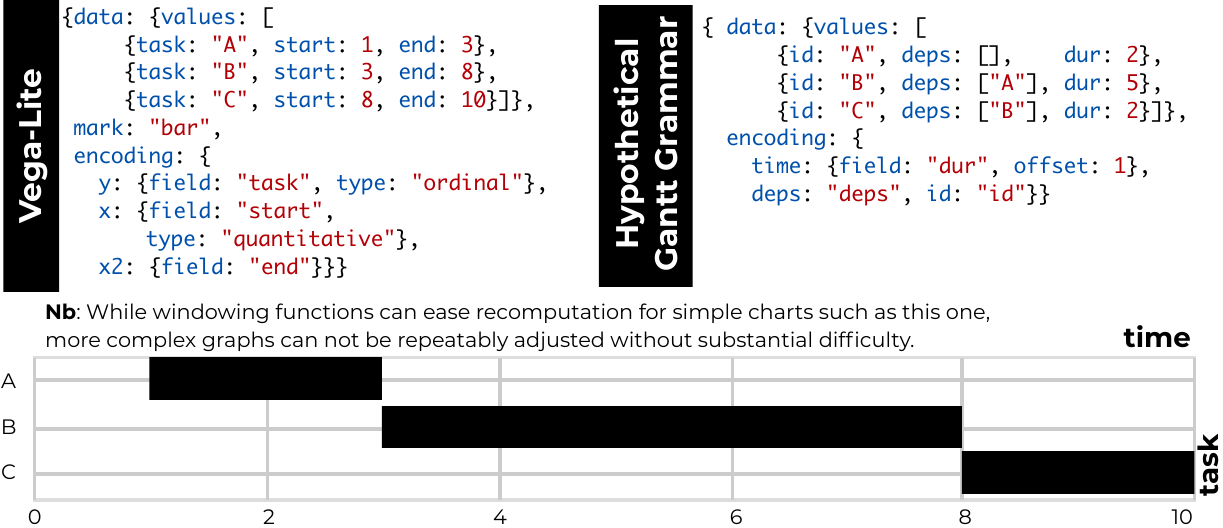}
  \caption{
    While these Gantt chart specifications are similar in length and complexity, the task-specific DSL does not require the user to manually update the positions after a data update.
  }
  \label{fig:gantt-chart}
  \vspace{-0.2in}
\end{figure}

Some chart families---like Gantt charts, Sunbursts, HOP Plots, or Euler diagrams---can be constructed through visualization DSLs whose notations are not well aligned with those charts.
For instance, while Gantt charts can be produced using general charting tools like \vegalite{} (\figref{fig:gantt-chart}), they do so in a fashion that is not well matched with the data. Gantt charts show the relationship between projects over time in a DAG structure, sometimes featuring graph-based computations such as \emph{critical path} (maximum path length), neither of which are well supported in the tabular data model used in many DSLs.
The result is that the user potentially needs to make tedious layout re-computations and round-trips to the data.
While this data type and associated transformations could be embedded into \vegalite{}, doing so would add to the growing complexity of that language.

We suggest that the creation of task-specific DSLs (or what Guzdial and Shreiner\cite{guzdial2021integrating} would call Teaspoon languages) that allow end-users to author charts in the vernacular of a particular chart or data form to be an important opportunity.
\encodable{} addresses this problem through a \vegalite{}-inspired component abstraction over arbitrary chart forms---although this is done in such a way that binds the charts to the JS-implementation (precluding portability) and requires a tabular data structure (which can be at odds with the domain, as in Gantt charts).

\parahead{Interactions.} Most languages in our survey produced interactive visualizations that supported at least some simple interactions, such as tooltips. However, some languages specifically focus on nuanced or uncommon interactions.
Some DSLs center animation, such as \anivis{}, \sys{Gemini 1/2}, and \canis{}.
\cicero{}, \echarts{}, and \zingchart{} emphasize creating responsive visualizations.
\svl{} supports some collaborative visual analytics interactions.
Focusing on a particular interaction allows the language to surface attributes that are specific to that interaction, such as \geminitwo{}'s use of keyframes as a first-class element of the language.

Careful balancing of novel and expected features is essential, as having too many options can dilute the specificity of the DSL, however, missing anticipated features can ruin its domain utility. For instance, zooming is an interesting addition to abstract visualizations in \gosling{}, while in map-focused DSLs it is all but required.

A number of DSLs focus only on the interaction of concern and offload the remainder of the work to another DSL through compilation (\secref{sec:languages}).
This style of interaction injection is a valuable component of some JSON-style DSLs. The restricted grammars allow simple language composition, allowing each DSL to do one thing well. Similar effects can be achieved with plugin architectures, however such methods are usually inaccessible to end-users.

Nearly all interactions in these DSLs are \emph{transient}, with only one DSL (\glinda{}) supporting updates from their state into their specification.
While some interfaces (\eg{} B2\cite{wu2020B2}) provide mechanisms to reify interactions into code, it is a missed opportunity that such \emph{bidirectional interactions} are relegated to external tools given JSON-like DSLs computational malleability.
For instance, tasks like annotation may be easier to complete using direct manipulation as it does not require repeated round trips to code\cite{victor_drawing_2013}.
We suggest that such tasks may be well-supported by DSLs that allow alteration of their specification from their output \emph{in addition} to from the code, such that changes to the output are reflected in the input and vice-versa.
Similar techniques have been employed to allow bidirectional updates to SVG drawings formed through functional programming languages \cite{hempel_sketch-n-sketch_2019, Glisp}.
As JSON-style DSLs can be manipulated more easily than most GPLs, we suggest that future systems should explore this intriguing feature domain.

\parahead{Mediums.} The purpose of a number of systems is to enable use of a particular medium, form of data, or set of tasks.
As noted in \figref{fig:family-table}, these DSLs address a wide range of domains including SciVis, big data, and data science/ML.
This broad purpose highlights that visualization tasks occur outside of the tidy small-data abstract-charting sandbox that visualization DSLs (and systems\cite{satyanarayan_critical_2020}) often target, and moreover, that conceptual adaptations can be usefully made to serve other use cases.

Focusing on domain allows selection of appropriate notation.
For instance, some domains use their own coordinate systems either by convention or necessity.
\genomespy{} and \gosling{} include an idea of genomic coordinates.
\comicscript{} has a comic-specific notion of panels.
While generic approaches could achieve similar ends, they would not match the expressivity found by localizing the syntax to the domain. \comicscript{} has limited support for data exploration but enables interactive data comics in a way that would otherwise be unmanageable\cite{wang2022interactive}.
As with DSLs in any domain, this may come at the cost of generality.

Surfacing domain-specific concepts allows the user to avoid the Turing tarpit and directly address aspects relevant to the domain. However, JSON-style DSLs are most useful for tasks that specifically fit the medium or that an end-user would wish to accomplish as a DSL (or through a facade for one).
For instance, several virtual or extended reality systems (XR)---such as \dxr{}, \vrviz{}---use JSON-based grammars as the basis of their syntax.
We suggest the prevalence of this approach may be because of XR's need to pass between mediums (\eg{} between JS and Unity) is well matched with the portability of JSON-style DSLs.
Tasks with less structure may be better matched with free-text DSLs (\eg{} ViSlang\cite{rautek2014vislang}) as these allow for greater flexibility at the expense of automation (\eg{} GUIs or recommenders).

\subsection{Models: What is it?}\label{sec:models}

All languages are predicated on a model of what computation is executed based on the commands written by the programmer.
DSLs in our survey used models that fell along axes of low to high abstraction, and formal to colloquial.
Drawing on prior work\cite{wongsuphasawat2020encodable}, abstraction level refers to DSLs that are close to the data domain as ``high-level'' (\eg{} \vizml{} or \highcharts{}) and those near the rendering context as ``low-level'' (\eg{}  \vega{} or \deckgljson{}{}).
Model formality denotes a pervasive logical structure in the DSL's design.

\parahead{Formal Models.} Languages backed by overarching frameworks are an important approach to visualization system design and are successful in tools like Tableau and ggplot.
These \emph{formal} models can simplify the expression of intent (within their scope) and aid potentially difficult analyses, although possibly impeding flexibility.
We observed a variety of models whose purpose and intent varied.

The most prominent of these is Wilkinson's GoG\cite{wilkinson2012grammar} model in which data attributes are mapped to encoding channels and combined through marks.
These forms allow for expressive construction and combination of visualizations allowing for the fluid creation of novel forms without concern for chart type.
Examples of this form include \vegalite{}, \vega{}, \vizml{}, \jsol{}, and \flexer{}.
We suggest that this form is a safe default model for many visualization systems, as it ``expose(s) the mechanics of good practice''\cite{healy2014data}.
Wilkinson has argued\cite{wilkinson2012grammar, pu2021special} that his Grammar of Graphics is the \emph{only} grammar of graphics.
While this framing has been enormously successful in the development of visualization systems and languages, it is far from the only conceivable systematic model for creating visualizations. Others can be more tightly tuned to support particular tasks.
For instance, \vizql{} emphasizes data exploration through the language of a data cube.

\begin{figure}[t]
  \centering
  \includegraphics[width=\linewidth]{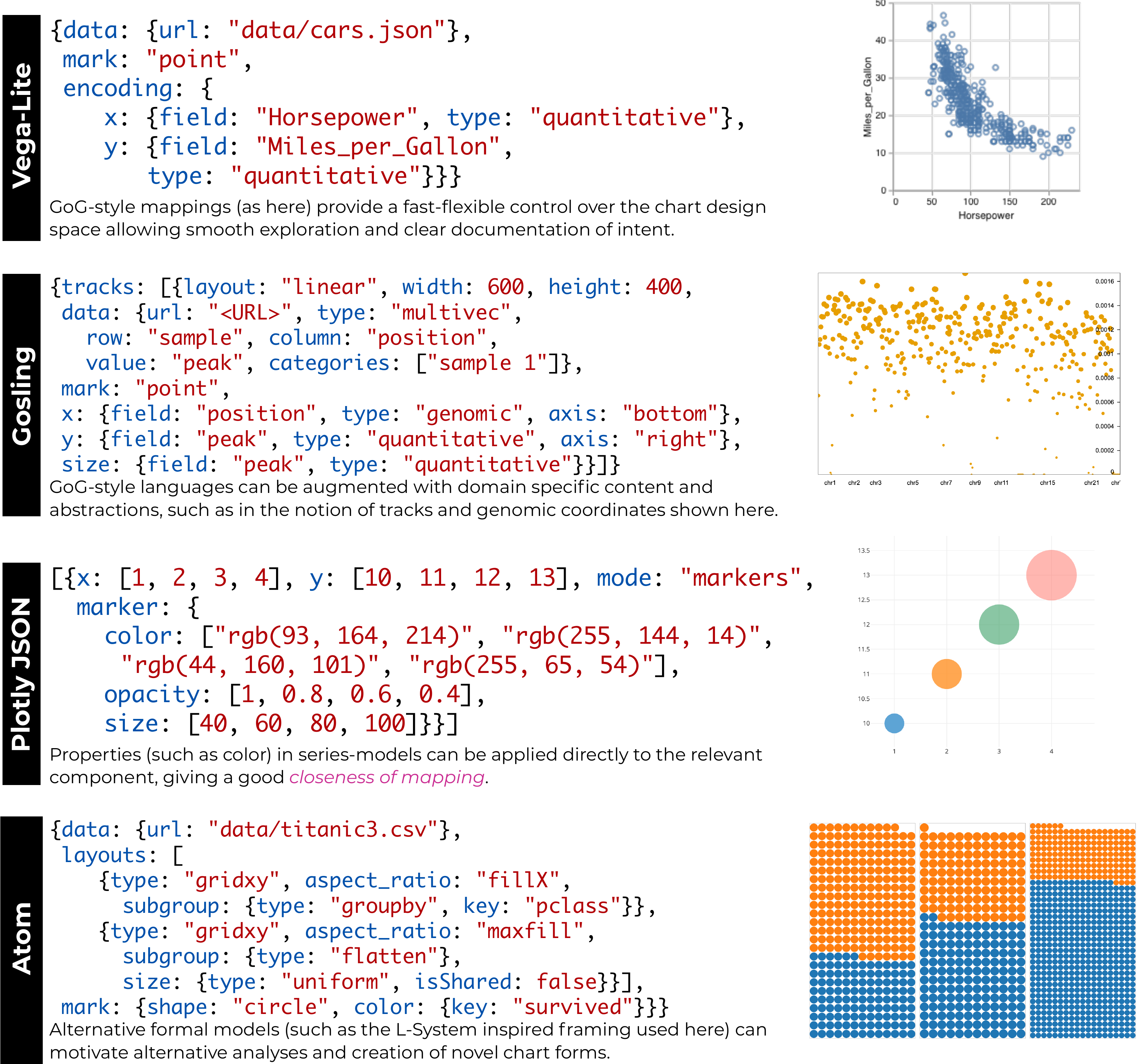}
  \caption{
    A DSL's domain and model manifest themselves in its syntax.
  }
  \label{fig:language-examples}
  \vspace{-0.2in}
\end{figure}

While many DSLs expressively use a declarative mapping of data to visual properties, other formulations can also be effective.
For instance, \atom{} uses an L-system-inspired model to describe unit visualizations (see \figref{fig:language-examples}), whose graphical forms carry pivot sequences that are well matched with the iterative stages of an L-system.
\compasql{} and \setcola{} use constraints to generate programs that can act as a facade for complex  systems whose interface might not be easy to understand or write (such as recommendation systems).
\ScholzthreeDVisLanguage{}, \ShihVolumeVisLanguage{}, and \sys{P4-6} use a sequential pipeline model.
Pipeline models seem to be especially suited to low-level graphics tasks (as in \ShihVolumeVisLanguage{} or shaders) or data manipulation (as in \pv{} or \vega{}'s data model) as they may involve iterated stages whose results must be produced sequentially.
We highlight these alternative framings because they may yield new approaches to various problems and enable chart types or analyses which are impossible (or needlessly difficult) in other framings.
Alternative conceptual models are not unique to our survey. For instance, the HiVE notation \cite{slingsby2009configuring} eases exploration of hierarchical orderings in treemaps.
The development of DSLs for particular chart types or aspects is an intriguing opportunity for this genre of work, as are DSLs that surface novel semantic models as a way to expose new framings for analysis.

While our survey criterion requires that every DSL have algebraic qualities, several DSLs introduced explicit algebraic structures.
Both \vizml{} and \vizql{} use table algebras  to describe the way  data is manipulated prior to graphical display.
\vegalite{} uses a simple data table model but provides a view algebra that provides affordances for viewing permutations and combinations.
These algebras allow for rich expression within their domains. Similar systematic models might be established to serve other tasks or simply to surface alternate analysis approaches. For instance, \neo{} uses an algebra specific to confusion matrices.
The visualization process has many interrelated steps, many of which might be enriched through formal modeling.

\parahead{Colloquial Models.}
The complement to formal models are those models that do not impose a framework over the structure of the interface.
These \emph{colloquial models} are sometimes overlooked despite their ability to adapt and accommodate real-world situations and problems.

These less structured DSLs typically utilize series-based models in which
data is mapped to aspects of \emph{particular} chart forms (as in templates), such as the x-axis of a scatterplot or the angles of a pie chart (\eg{} \plotly{} in \figref{fig:language-examples}).
Systems whose specification unit is a layer (such as \deckgljson{} and \bertin{}) can be seen as using a series model in which the series are superimposed.
They can be used at any level of abstraction, such as high-level DSLs like \anivis{} (which focuses on high-level chart templates) or low-level DSLs as in \deckgljson{} (which can give access to shader-level manipulations).

\begin{figure}[t]
  \centering
  \includegraphics[width=1\linewidth]{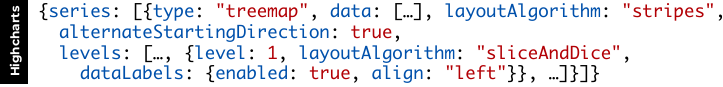}
  \caption{Colloquial DSLs can enable specification of chart attributes without modeling those attributes through the entire system.}
  \label{fig:informal-model}
  \vspace{-0.2in}
\end{figure}

While these approaches are sometimes denigrated for their perceived lack of expressivity \cite{wilkinson2012grammar, Friendly22Colorless},
the close connection between their inputs and outputs forms a \cogdim{closeness of mapping} that can make them simple to understand and easy to verbally describe (aspects which formal approaches may fall short on).
This closeness can make the process of switching to a conceptually related (but visually distinct) chart form more \cogdim{viscous}, as the two syntaxes may be highly different.
They are more likely to allow unusual graphics as they can be created without respect to a formal model, such as \bertin{}'s  multiple forms of cartograms or \zingchart{}'s funnel charts.
These interfaces can be tuned to specifically support the domain, as in \highcharts{}'s treemaps (\figref{fig:informal-model}). However, their design is often ad hoc and may be \cogdim{inconsistent} with other parts of the system, rendering them harder to learn or understand.
This approach allows for simple local styles to be applied without needing to map those properties through top-level language concerns, as well as hierarchical styling which can be difficult to apply in systems whose data model requires tabular normalization for rendering (as in \vega{}).

Colloquial models can support particular features without forming themselves around that concept.
For instance, only a handful of DSLs have top-level annotation support (\eg{} \plotly{}, \highcharts{}, and \apexcharts{}), however most of these are industry-driven efforts that do not use a formal model---\pfour{} and \cicero{} excepted, as they explicitly model annotation.
Similarly, only a small set of mostly industry-led DSLs provide their own accessibility features (\highcharts{}, \vegalite{}, \echarts{}, and \fusioncharts{}).
This dearth may be due to the fact that such practicalities are often viewed as mere implementation details rather than being central to usability\cite{tsandilas_vis2021, park2017atom},
which may be compounded by a lack of research incentives to provide usable artifacts.

Beyond explicitly modeling features, formal models can provide these model-breaking behaviors through \emph{escape hatches} to non-declarative programming or other points of extension (\secref{sec:practice}). However, doing so requires that such hatches be pre-placed in a manner relevant to the new feature---which is difficult to predict.
Exploration of DSL malleability in a manner that allows for undesigned features without becoming unapproachably complex for end-users (a concern for the deeply malleable Varv\cite{Varv22Borowski}) is a valuable opportunity.

Formal models may have colloquial components. These are often found at interfaces with other systems.
Styling is one such common leaky abstraction. The manner in which a chart is rendered may leak into the formalism without being modeled.
For example, \vegalite{}'s concept of styling is guided by its downstream SVG and canvas renderers.
Colloquial DSL components are not inherently detrimental. If the leaked feature is large or complex, it may be appropriate to  embed that language rather than model it, although this can lead to \cogdim{inconsistent} interfaces that are hard to adapt to new forms.
Consideration of these properties in DSL design may be valuable, such as by designing with leaking in mind or by separating external concerns into separate DSLs.

Despite their academic stigma, we suggest that colloquial models may be valuable to consider.
Among series-based DSLs we observed 306 distinctly named series types, although there was significant overlap in this list  due to synonyms or simple modifiers (\eg{} ``3D'' or ``drag-able'').
The design space of name and modifier-based specification appears to be a rich one,
although an analysis of which is beyond the scope of this work.
Developing a better understanding of these forms and the way they are used by system designers and domain experts is intriguing work---especially in light of the decades-long popularity of this specification style (\cf{}  SpotFire, \ChartML{}, \xmlcharts{}).
As researchers and tool-smiths\cite{brooks1996computer}, we suggest that we should meet people where they are, which may mean working to enrich colloquial models.

\newcommand{\compilation}{}
\newcommand{\extending}{}
\newcommand{\embedding}{}

\newcommand{\influence}{$A$\hspace{0.02in}\faLightbulbO{}\hspace{0.02in}$B$}

\begin{figure}[t]
  \centering
  \includegraphics[width=\linewidth]{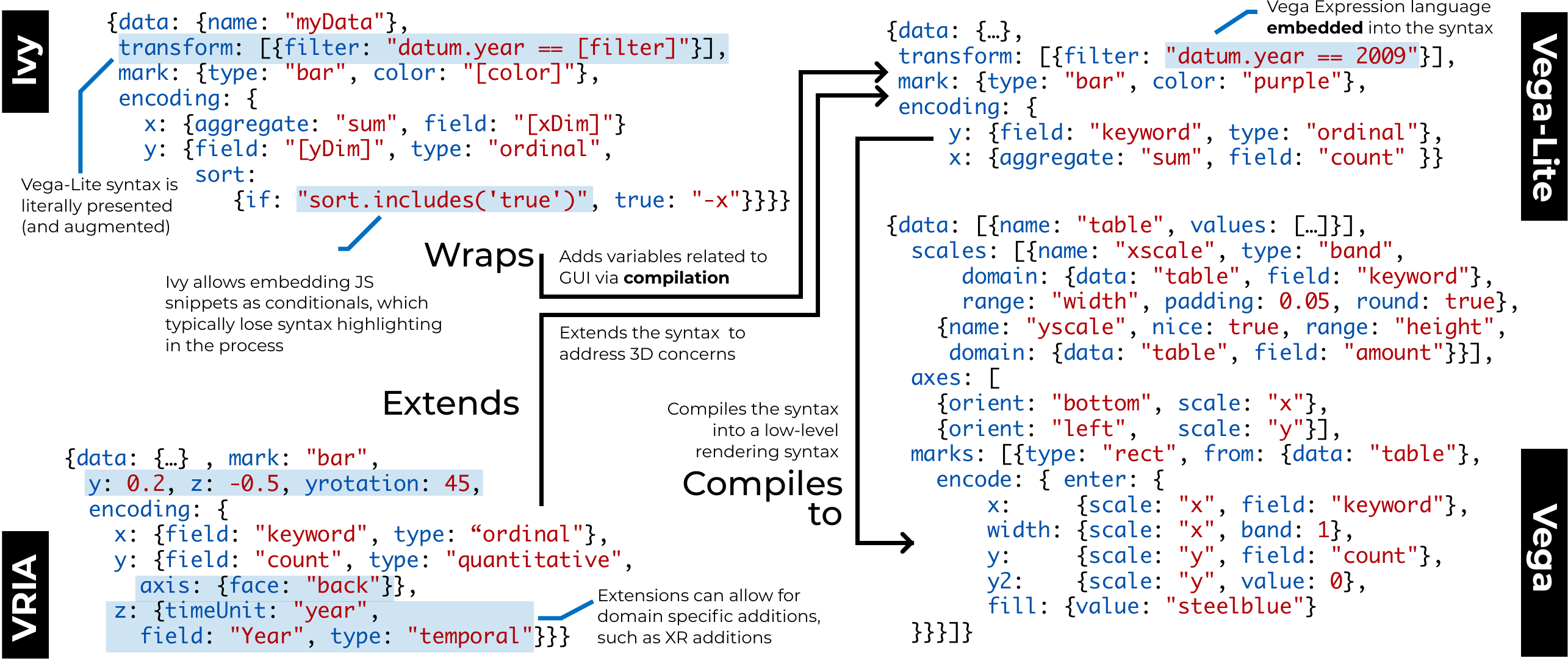}
  \caption{
    Annotated examples of the relationships between several DSLs.
  }
  \label{fig:rel-examples}
  \vspace{-0.2in}
\end{figure}

\subsection{Relationships: How does it relate to other DSLs?}\label{sec:languages}

Languages rarely exist in a language vacuum.
For instance, many languages support regular expressions in a syntax unrelated to their own.
We found that our DSLs are related to each other by Compiling\compilation{}, Wrapping/Embedding\embedding{}, or Extending/Contracting\extending{}.
The way a language relates to others determines numerous details about its implementation (particularly its execution model), although each relation has tradeoffs---the main tension residing between language customization and reuse.

Language composition can allow DSLs to be developed without requiring re-implementation or invention of abstractions.
Yet, this comes at a cost as the new system is limited by the design choices of the old.
For instance, compiling into \vegalite{} ensures that \cicero{} does not require the design of a visual encoding system, however, doing so means that it is restrained to the mark types available in \vegalite{} (and by a similar composition, \vega{}); impeding forms such as cartograms or treemaps.
While such tradeoffs can be navigated, compositions that surface nested DSLs to the end user, as in \ScholzthreeDVisLanguage{} use of \vega{} specifications, may diminish the value of a language style API,
As they may necessitate frequent reference to documentation of the nested-DSLs rather than providing a single coherent expression language---yielding \emph{language cacophony}.

\parahead{Compile\compilation{}.}
Among \emph{external} DSLs, there are two evaluation mechanisms: compilation and interpretation\cite{fowler2010domain}.
We refer to compilation as a process that generates code, while interpretation evaluates it directly.
\emph{Internal} languages are embedded into their host which carries with them all of the benefits and drawbacks associated with the more generic DSL design decision of internal vs external.

Compiled languages allow the language to gain all the strengths---and weaknesses---of its compile-target (often an interpreted or internal DSL), such as with \cicero{}.
Compilation does not require a direct translation but can be used to embed computations into the resulting system.
\setcola{} uses this strategy to create circular layouts which are not present in its compile-target  WebCoLa\cite{DwyerWebCola}.
In addition to offloading rendering, this gives access to potentially hard-to-achieve functionality, such as accessibility features.

Compilations may be chained together in a \emph{compile tower} in a manner supported by targeted languages that do one thing well (akin to the Unix credo). For instance, our Gantt chart example in \figref{fig:gantt-chart} might usefully target \vegalite{} to gain features such as tooltips without requiring implementation of nuanced details. As the ecosystem of JSON-style DSLs continues to grow it may be advantageous to select designs that re-use as much work of the previous DSLs as possible.

These benefits  do not come for free. While \vegalite{} provides ARIA-accessibility features, it currently cannot provide some accessibility-enhancing encodings (\eg{} texture) because its compile target, \vega{}, does not support them.
More generally, errors may be harder for the end-user to understand if they are generated by the target's interpreter, whose concerns and conceptual model may be different from the source language.
\ivy{}, which is a wrapping language that \emph{uses} compilation, exemplifies this duality. It is language-agnostic and can be used over any JSON language, however, doing so precludes the surrounding application from providing contextual hints because it is unaware of the languages over which it is executed.
Despite this, we argue that while compile towers are not always applicable, they should be employed more often. They can simplify implementation, improve usability, and reduce reinvention.
They are well suited to research systems (whose contribution is not based on implementation) as they can rely on another system for repeatability and defense against bit-rot.

In contrast, interpretation allows for rich customization that can be helpful in specialized contexts.
\kyrixs{} supports large data sets for zooming visualizations. \DataTheater{} uses an unusual data model (the output of an end-user specified Python script) to create explorable explanations.
This approach can enable construction of contextual error messages (and other usability features) that are relevant to the local domain as they are not predicated on layers of indirection.
However, this approach pushes rendering, data manipulation, and usability features onto the language implementer.
Medium-focused DSLs tend to use this approach, possibly because their value is related to their customization to that medium (the main exceptions to this are XR-focused DSLs).

\parahead{Wrap/Embed\embedding{}.}
Wrapping or embedding languages provides functionality extensions by literally containing other languages.
For instance, \geminitwo{}, allows users to describe keyframes of an animation by explicitly including \vega{} and \vegalite{} specifications.
\ScholzthreeDVisLanguage{} allows the inclusion of entire \vega{} and \vegalite{} charts in a 3D context.
There is overlap with compilation (as it can be used as a wrapping mechanism, as with \ivy{}), however we delineate this as a separate pattern to highlight the particular form of re-use.
This approach allows for language-level separation of concerns as well as the use of the imported DSL's externalities (\eg{} documentation or community support). This approach's main risk (beyond language cacophony) is that the embedded language might not match the domain and lead to  \cogdim{inconsistencies}.

A less extreme example of this approach is to embed language snippets---such as in the manner that SQL snippets are represented as strings in GPLs.
These snippets address common tasks, such as control-flow or formatting (often via the d3-format language) as well as model-specific issues.
\vega{} has a purpose-built JS subset for interacting with event streams.
\fusioncharts{} permits HTML snippets in tooltips.
This common DSL pattern\cite{fowler2010domain} allows for rich expression of intent, but may come at the cost of tooling,
yielding some usability features (\eg{} syntax highlighting) unavailable.
There may be \cogdim{hidden dependencies} within the snippet, as in the often numerous signals in \vega{} expressions \cite{hoffswell2018augmenting}.

\begin{figure}[t]
  \centering
  \includegraphics[width=\linewidth]{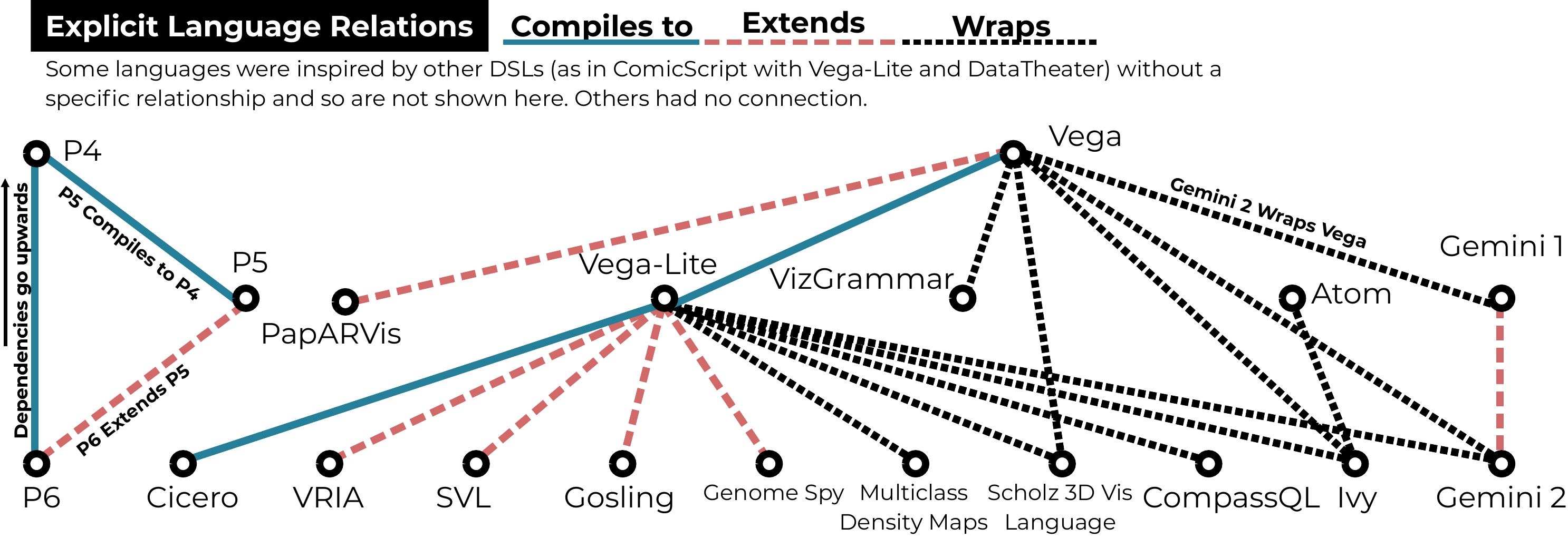}
  \caption{
    DSLs can hold a variety of relationships with one another that allow them to reuse implementations, syntax, or concepts.
  }
  \label{fig:family-tree}
  \vspace{-0.2in}
\end{figure}

\parahead{Extend/Contract.\extending{}}
An associated relationship is extension, in which a DSL is contracted or extended to form a new DSL.
This usually comes in conjunction with syntactic extensions or modifications to the execution strategy.
\genomespy{} and \vriajs{} contract the syntax of \vegalite{} behind custom renderers, and extend it with some genome and XR-specific affordances, respectively.
\PapARVis{} wraps and extends \vega{} with augmented reality enhancements.
This approach can be useful as it allows for porting of ideas to new domains
(\eg{} \genomespy{}'s reuse of \vegalite{} syntax in genomics).
However, it does so at the expense of creating a new backend for that system.
Given the variety of functionality developed across these DSLs,
enabling their composition to allow greater reuse and increase their long-term impact.

Language elements are sometimes extended or reused in an ad hoc manner, a pattern which is more closely aligned with influence than extension.
Some languages (such as \encodable{}, \flexer{}, or \dxr{}) explicitly mold themselves on the thin mapping style of \vegalite{} or \vega{} without actually reusing the specific syntax or rendering systems.
Other DSLs include only minor syntactic elements, such as \pfour{} and \ivy{}'s use of MongoDB-style operators (\eg{} \inlineFig{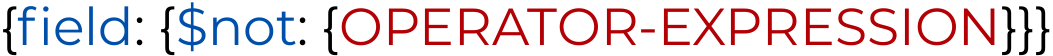}).
At other times this influence is conceptual. \gotree{} includes a spacing system related to the CSS box model, while \zingchart{} and \cicero{} include responsive-design features inspired by CSS media queries.
Familiar syntax and concepts may aid learnability---possibly reducing the negative effects of \emph{language cacophony}---however not every domain will fit every imported idea.
For instance, CSS-style declarative rules are unlikely to handle the iterative stages of a data transformation pipeline well.

\subsection{Language Affordances: Who is the end-user?}\label{sec:affordances}

Each language has at least a general idea of its users, which motivates what features to include.
We saw three user types based on the way they are expected to use the DSL, whose interests are naturally in tension.
Some users simply use the DSL (\emph{end-users}), others can modify the system which houses the DSL (\emph{system-builders}), while others automatically manipulate and analyze the DSL (\emph{automated agents}).

\parahead{Syntax.} A common first choice is whether to create an internal or external DSL.
This is covered at length in other venues\cite{fowler2010domain,mernik2005and} but in essence, it can be seen as a question of language invention or embedding.
The languages in our survey do not demonstrate any substantial divergences from the common benefits and limitations of each of these patterns. External DSLs offer richer expressivity but can be harder to construct and learn. Internal DSLs are easier to use within a host language but can force a notation that is poorly matched with the domain.

Internal languages seem to be most useful to interface-builders\cite{wongsuphasawat2020encodable} as opposed to end-users.
For instance, they do not by default facilitate automated analysis, and as such analysis can require dedicated high-complexity tools (such as AST-analyzers).
These components are often beyond the design goals of tools meant only to support web-based presentations (\eg{} \chartjs{}).
Notebook-based analysis is a notable exception as it blurs end-user and system-builder.
Internal bindings to external DSLs (\eg{} Altair\cite{vanderplas_altair_2018}) seem well matched to such hybrid users.
However, such an analysis is beyond the scope of our study.

Among external languages, JSON is sometimes chosen for being end-user friendly. Scholz\cite{scholz2021modular} notes that JSON was selected because it is human-readable and easy to transfer on the web.
DeLine valued ``YAML's declarative, hierarchical syntax''\cite{deline2021glinda}.
In contrast, a common criticism of XML as a carrier language is that it is verbose, which is seen as poorly matched with human usage \cite{fowler2010domain,wilkinson2012grammar}. JSON's syntax, which is \cogdim{terse}, appears to overcome this hurdle and may account for this style of DSL's growing popularity (\figref{fig:over-time}).

Several end-user-focused systems expose their syntax to the end-user in applications. \structgraphics{} uses a custom GUI, whereas others (\eg{} \ivy{} and \glinda{}) use plain text extended by modern editor affordances (\eg{} autocomplete). Other DSLs have editors that support their use (without being required), such as \sys{Gemini 1-2} or \vega{}.
Constructing an environment around a DSL allows debugging tools and other end-user supportive features, however, this can (and has\cite{wu2014case}) led to a constellation of small applications that repeatedly reimplement similar functionality. We suggest that it may be beneficial to consider how these efforts may be consolidated for this language style more generically.

Shih \etals{} \cite{shih2018declarative} rationale for selecting JSON for their scientific visualization grammar was more focused on machine usability, noting that they selected ``JSON because it is a widely used standard, is easy to parse, and it has sufficient expressiveness for hierarchical structures'', an attitude shared in the design of \comboCite{\gotree{}}{\cite{li2020gotree}}.
Wu \etal{}\cite{wu2021ai4vis} note that this interface style allows for manipulation by humans and autonomous users.
While true for any executable language, manipulations are easier in serialization formats due to their restricted form.
The limited grammar allows for exhaustive design space exploration, enabling recommendation (as in \compasql{} and \cicero{}\cite{Hyeok22Cicero}) and enumeration of novel chart forms (as in \comboCite{\atom{}}{\cite{park2017atom}} and \comboCite{\gotree{}}{\cite{li2020gotree}}).

An often discussed benefit of external DSLs\cite{mernik2005and} is that they expose a notation local to a domain---as in Diderot's\cite{kindlmann2015diderot} explicit use of tensor operators (e.g. $\nabla$ and $\circledast$).
While JSON-like languages can abstract over various domains, few domains use it as their primary notation (API design and data definition are clear exceptions).
The selection of these carrier languages as syntax is then a compromise. In exchange for benefits like portability and simple machine operability, domain experts encounter a less familiar notation.

\begin{figure}[t]
  \centering
  \includegraphics[width=\linewidth]{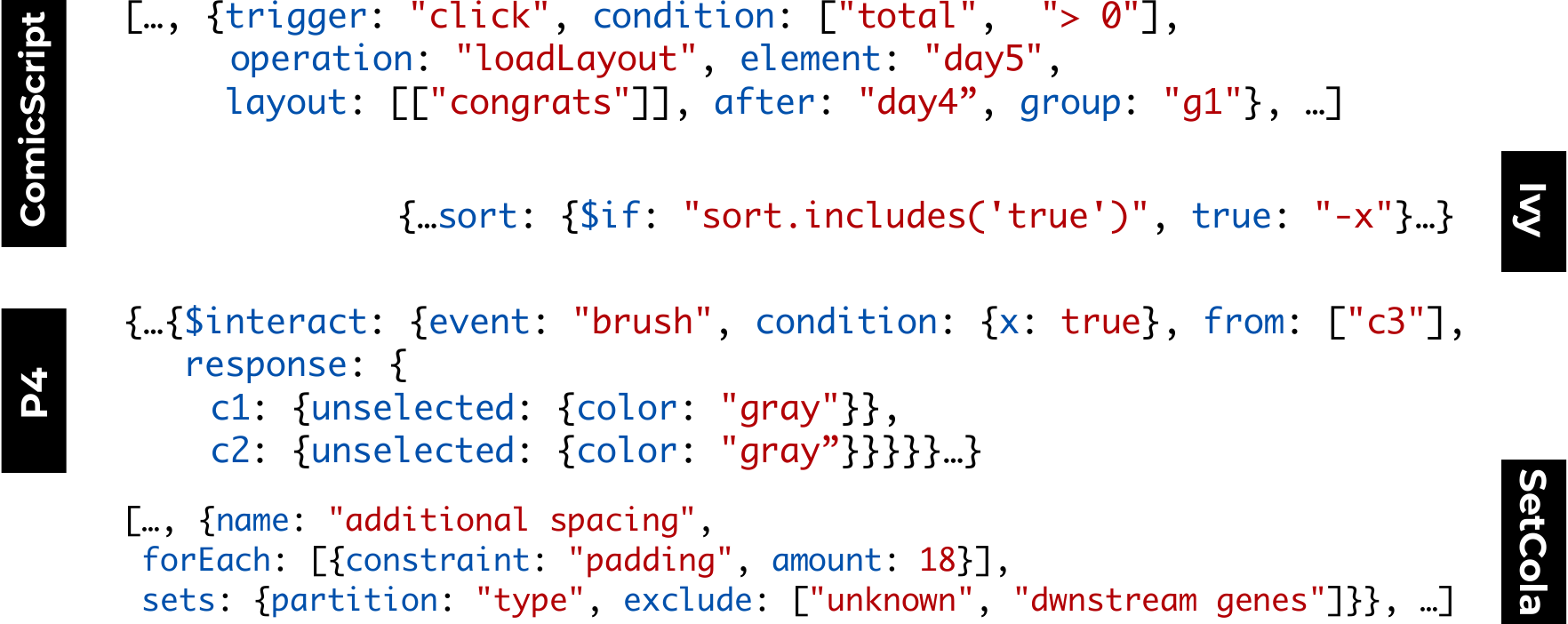}
  \caption{
    Many languages feature logic or control flow operators.
  }
  \label{fig:logic-ops}
  \vspace{-0.2in}
\end{figure}

\parahead{Abstraction Mechanisms.}
Creating \cogdim{abstractions} is an important part of any programming language.
In GPLs features like variables, functions, if-else structures, loops, and a host of others serve this purpose.
Some languages in our survey utilized these elements (\figref{fig:family-tree}) allowing abstraction on syntactic, data, output, or contextual levels.
Those that did not, likely did so to limit scope, because their domain did not require it, or relied on their host for such features (a benefit of internal DSLs).
The tendency to forgo abstraction in DSLs is well known\cite{fowler2010domain}, but we highlight it to explore the particularities exemplified in this context.

Control flow operators (as in \figref{fig:logic-ops}) were common.
These conditionals can address a range of program aspects including the data (as in \vegalite{}'s conditional marks), the graphic (as in the query selectors found in \canis{}), interactions (as in \comicscript{}), or container state (such as in languages rooted in GUIs like \ivy{}).
Some focused on modifying graphics based on interactions (as in \pfour{}) while others focused on syntactic transformations (as in \ivy{}).
Some languages, such as \vega{}, use another language to evaluate their condition (via embedded snippets), while others, like \comicscript{}, construct the logic through explicit operators.
While it is not necessary to be able to query or conditionalize every element, each of them can be beneficial depending on the domain although not every situation necessitates such facilities.
\cicero{} uses a powerful query language that gives access to data, graphic, and specification, which is necessary for its responsive and annotation tasks, although it appropriately has no concept of its surrounding context (besides aspect ratio).
Embedding snippets offers greater expressivity\cite{fowler2010domain} (potentially at the price of portability and diminished usability features), while explicit operators allow the user and their tooling to keep a single consistent mental model (potentially introducing unfamiliar syntax).

A variety of other abstraction mechanisms were used.
Some systems included notions of variables, although their purpose varied.
\flexer{} and \vega{} use FRP-style signals as variables.
\canis{}
uses variables as a form of textual-macro replacement.
\ivy{} and \vegalite{} use variables as a way to reference GUI controls exposed to the end-user (although \vegalite{}'s are a mask for \vega{}'s signals).
Variables can help reduce the cognitive load on the user by reducing \cogdim{diffuseness}, but it can also increase it if the references become difficult to follow.
\setcola{} was the only language to include loops. While an appropriate syntactic choice for their domain (simplifying constraint generation) it is common for DSLs to not provide loops as this can cause DSLs to accidentally ``slide into generality''\cite{fowler2010domain}.
None of the \emph{external} DSLs had SQL-style end-user definable functions.
Varv\cite{Varv22Borowski} takes extensibility to an extreme via a fully end-user editable application creation external DSL (that includes simple macros); demonstrating that this level of malleability is achievable in external DSLs.

The selected \cogdim{abstraction gradient} should cater to a designed audience.
Loops and variables can help readability, which supports humans but does not generally affect automations. Functions and control flow operators can aid in reuse, but if programs are generated on the fly in a GUI and not meant for reuse, their utility will be limited.
If the intended user's interests are not aligned with multiple such user types then a different interface may be preferential to a JSON-style DSL.
For instance, a DSL solely focused on humans using notebooks will likely be better served by not imposing the grammatical limitations of a serialization language, while an automatically generated language for facilitating chart recommendations need not be human readable.

\subsection{Practicalities: Where is work done?}\label{sec:practice}

\begin{figure}[t]
  \centering
  \includegraphics[width=\linewidth]{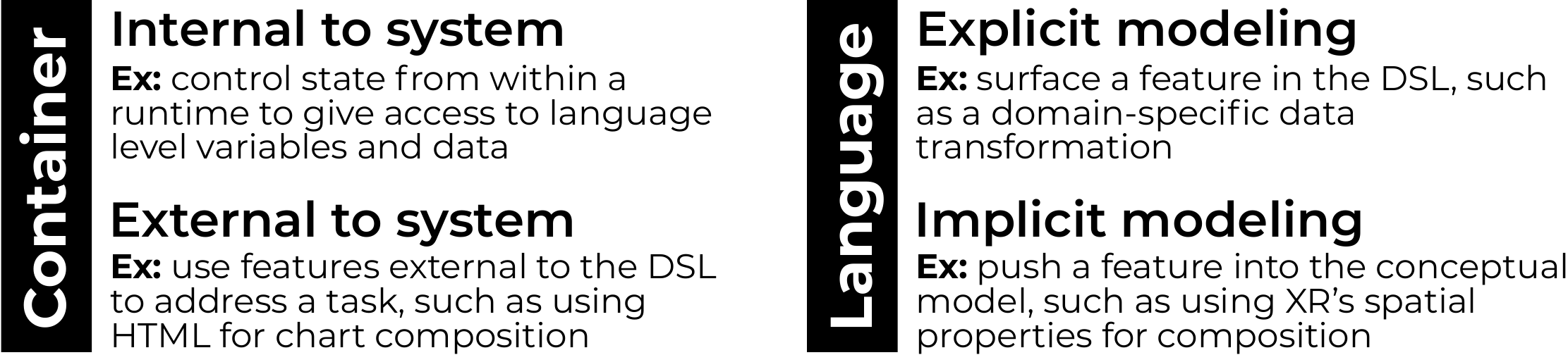}
  \vspace{-0.2in}
  \caption{
    Features can be built in a variety of places across the DSL.
  }
  \label{fig:work}
  \vspace{-0.2in}
\end{figure}

There are a number of places within a DSL where a given feature can be implemented.
As in \figref{fig:work}, these include explicit and implicit modeling as well as internal and external placement.
This modeling describes where the user is expected to do work to use those features.

Each strategy has advantages and disadvantages.
Internal features give deep control over implementation. However, that entity must be clearly represented or risk \cogdim{visibility} errors.
Explicit modeling can allow the user to address a task directly, but it can require the development of new (potentially \cogdim{inconsistent}) syntax.
External features can push burdens to other systems, reducing portability and potentially inducing \cogdim{hidden dependencies}.
Implicit modeling can be deeply expressive, but carrying out such intentions can yield \cogdim{hard mental operations}.

\parahead{Integration and State.}
Most DSLs manage state and interactivity through a runtime inside the system.
This allows them exact control over the way a feature is delivered, and is thus favored by internal and interpreted DSLs.
Similarly, internal DSLs allow rich integration with web pages through affordances like callbacks---typically in exchange for a lack of end-user control.
Other systems manage state by integration with an external application.
\structgraphics{}, for instance, provides a visualization builder interface that maps spreadsheet data to graphics.
Embedding state into a housing application allows graphics to be synchronized with and used to control the UI, enabling deeply integrated experiences, although this may impede portability.
Some systems (typically compiled DSLs) pass control to an external system, as in \vegalite{} or \setcola{}.
This simplifies system construction, however, it may impede interactions outside the target's model.

These approaches only have value when appropriately coordinated with their purpose.
For instance, \atom{} uses a custom interpreter integrated into an application.
While this approach allows \atom{} to be closely integrated with its editing environment, it makes it difficult to \emph{portably} reuse specifications in other contexts and precludes them from being integrated into other applications.
Given its position as an academic artifact whose value is not related to its renderer, we suggest that DSLs like \atom{} may be well matched with a \emph{compile tower}-style strategy to alleviate implementation burden and facilitate portability.

Some DSLs provide mechanisms for interactions and integration with their environment, although this can require coupling with those systems (as does binding to any external system).
For instance, \encodable{} explores creating \vegalite{}-style facades over arbitrary JS visualizations. However, this causes those little languages to be inextricably linked to JS.
We suggest that \emph{portability and contextual integration are opposing goals}, as surfacing integrations as first-class aspects of the language creates a context dependence.
Both are reasonable design choices, but favoring integration may reduce some of JSON-style DSLs' utility (\eg{} portability).
Yet, language-level integration in these DSLs is unexplored, so it may be useful to consider visualizations as part of a system rather than singular units.

\parahead{Alternate language APIs.}
Some DSLs are used in host languages through internal bindings.
This type of tool can enable work-arounds for DSL limitations by externalizing these needs to a host language, as in the Gos Python-wrapper for \gosling{} \cite{l2022gosling} or Altair\cite{vanderplas_altair_2018} for \vegalite{}.
For instance, some facet and layer combinations create data ambiguities that can prevent \vegalite{} from rendering.
This can be resolved by manually pivoting data and combining Altair charts in Python.

Beyond providing workarounds for language issues, this can simplify program specification for users with limited familiarity with the DSL.
For instance, it may be easier for an Elm programmer to use elm-vega\cite{wood_design_2019}  than to a write a corresponding \vega{} specification.
This is analogous to the value of object relational mappers for manipulation of SQL databases. Users can write and reason about their database in their chosen language rather than being required to utilize SQL's sometimes idiosyncratic or unfamiliar form.
We emphasize that consideration of the environments in which a language will be used is a valuable component of language design as it may surface components and strategies that guide the design, such as expecting faceting to be done outside of the DSL.

\parahead{Extensibility.} Wilkinson notes that any closed system will have missing pieces\cite{wilkinson2012grammar}.
For instance, his \vizml{} has limited support for nested or data-driven layouts (\eg{} sets or cartograms) or mixed data and aesthetic-driven tasks (such as annotation).
Some DSLs approach extensibility by designing places where external elements can be introduced into the system through an API.
This can allow for the introduction of new transforms (as in \vegalite{}), user-defined marks (as in \dxr{}), renderers (as in \vega{}), as well as chart types or events (as in \chartjs{}).
Extensions typically occur in explicitly modeled features within the system, precluding end-user modification of system concepts (\eg{} new coordinate systems).
Open source software can allow for a slower but less limited form of extension. Yet this is not always the case. Some systems are no longer maintained, might be resistant to external changes, or might require too high a technical barrier to contribution from domain experts.

A form of extension available in some formal model-based DSLs is the creation of ad hoc mark types within the language itself.
Wongsuphasawat\cite{WongsuphasawatNavigating2020} explores how end-users of some systems can construct composite marks, such as candlesticks, although there are boundaries to this imposed by the form of the language model.
For instance, \vegalite{} allows some custom glyph creation (via image marks), which enables unit isotypes but not aggregates---as such encodings fall outside its model.
Some extensions are not possible without external modeling, as in our Gantt example (\figref{fig:gantt-chart}).
This extension style is powerful but is limited by its model and so can be well paired with external extension.

\parahead{Combination and Data Strategy.}
A DSL's approach to image composition (\eg{} layering or juxtaposition) and handling data exemplify the stratification of \emph{where} work is done.

We observed a spectrum of strategies for image composition ranging from specification \emph{above} the language (in the container), explicitly modeling \emph{within} the language, to \emph{below} it implicitly in the conceptual model.
Many DSLs do not provide a mechanism for combination, either because it is not relevant to their domain (as in graph DSLs) or by making use of awareness of their medium as the implied context through which conjunction happens.
For instance, \echarts{} relies on the browser for spatial arrangement and the user for data partitioning.
This can be simple to construct but pushes implementation onto the user.

Some formal models feature a composition algebra, as in \vegalite{}'s layer, facet, and concat operations.
These operators typically focus on data partitions (to facilitate small multiples), however, Wu also describes an under-explored notion of parameter-based faceting \cite{wu2014case}.
Some languages include a combination mechanism unique to their domain, such as \comicscript{}'s panels or \gosling{}'s notion of tracks.
These approaches are useful, but typically require explicit modeling in the language, which can take up limited conceptual real estate.

Some DSLs push feature description into their conceptual model, such as by using spatial position for combination (as in map and XR DSLs).
While powerful, these should be used cautiously as implicit operations can yield \cogdim{hard mental operations}.

The selection of how and where data is handled is critical, as it determines how a DSL can interact with its environment and compose with other DSLs.
Most DSLs hold all their data inside the system, a simplistic model which is adequate for many use cases, however some externalize that task.
For instance, \kyrixs{} uses a custom back-end that sits on top of a database to allow exploration of large datasets via zooming.
While sometimes useful, a complex data strategy is unlikely to work with a system that does not share that strategy: \kyrixs{} is unlikely to be interoperable with \MultiClassDensityMaps{} despite sharing a domain interest in aggregating heatmaps.
Similarly, DSLs exhibit no manipulation strategy (ignoring it or externalizing to the host), rudimentary language manipulations such as filters, or a richer domain-relevant expression or transformation system explicitly modeled within the language
(as in \genomespy{}'s genome-focused transformations).
These carry similar tradeoffs as composition in placement stratification. Unlike composition, there are well-implemented libraries that support this task.
Yet, many DSLs implement their data processing features themselves\cite{wu2014case}. These recreations may be motivated by domain. For example, most data libraries do not support the genomic coordinates required by \gosling{}---although this is more often done needlessly when more robust implementations are available.
Consideration of how and where to place features such as these is critical for DSL usability.

\section{Discussion}\label{sec:discussion}

In this study, we surveyed JSON-style DSLs for visualization from across academia, industry, and open source efforts spanning a period of more than 20 years.
In doing so we examined both the state of the art for this domain (such as the role of DSL compilation vs interpretation or abstraction in DSLs), and introduced new concerns (including the tension between colloquial and formal visualization models), patterns (like the role of composition), and practices (\eg{} supporting both computational and human users).
We observed a wide variety of tasks and domains that this style of language seeks to serve, indicating its pliability to a large collection of concerns and highlighted many avenues for future work (such as designing languages that can be bidirectionally updated).
From our results, we are optimistic about this style's future. However this is not without caveats nor enticing avenues of exploration.

\parahead{Study Limitations.}
We sought to understand the design and implementation of JSON-style DSLs through analysis of artifacts, documentation, and scholarly works. While this revealed a  number of intriguing patterns, it did not capture the entirety of visualization DSLs or APIs.
For instance, we excluded a variety of visualization languages (\eg{} ggplot) and JSON-based languages outside of visualization (\eg{} Varv). Exploration of design patterns and tradeoffs found in these and other languages, libraries, and APIs is warranted in future studies.

Computer languages, like spoken languages, are often living entities whose design changes over time and can be driven by individuals' undocumented ideas or influences. In future work, this analysis might be enriched through interviews with DSL authors to better understand language design choices and life cycles.

Our survey was biased in several ways.
Our survey was biased towards more recent open source and academic works, and away from older or privately-produced DSLs, as it is easier to find public contemporary systems.
As such, there are likely additional DSLs that were not observed during our search.
While additional data would be useful, we believe our sample sufficiently captures the tendencies of this language form, although sample size and biases are known issues\cite{zaytsev2015grammar}.
Some codings were based on limited documentation as we were unable to locate some DSL artifacts (\eg{} due to URL-rot or closed-source).
We intend to continue expanding the corpus of examples in our \asLink{\liveurltext}{supplement} to facilitate further empirical investigation of this language genre. This may reveal patterns hidden from our qualitative lens.
Finally, our analysis is limited by our own biases, which we sought to reduce through iterative theming and reflection.

\parahead{Language Design and Tooling.}
The design of an effective DSL for achieving any of the nuanced tasks that visualization DSLs seek to solve is a thorny problem.
Novel languages and notations have the potential to serve as foundations on which to ``think the unthinkable'' \cite{MediaForThinkingTheUnthinkable}, but also can add needless complexity and cacophony.
How to build powerful tools that do not result in confusion which can be applied to an ecosystem that has a competing set of users? Per the tensions and tradeoffs we highlight throughout this study, there is no one answer.
We suggest that developing mechanisms for design evaluation and improving the DSL tool ecosystem may fruitfully guide future DSLs.

DSL evaluation is a long-running topic\cite{poltronieri2021usability, borum2021designing}.
However, these methods are not extensively used for visualization DSLs\cite{pu2021special}.
Only two works\cite{satyanarayan2014declarative, devkota2021cfgconf} offered a formal CDN analysis, while others provided ad hoc reasoning \cite{wilkinson2012grammar}.
Poltronieri \etal{} \cite{poltronieri2021usability} suggest that DSL evaluation may be more effective if done in a contextual and non-generic manner (as in Jakubovic \etals{}\cite{jakubovictechnical} work on programming systems or Elavsky \etals{} work on visualization accessibility\cite{Elavsky22Chartability}).
Such a call might be answered with evaluatory heuristics, like
\emph{``What tasks does this DSL address?''},
% \emph{``does this language allow for details-on-demand style interactions?''}, 
\emph{``What form of model is it using?''},
\emph{``How are non-data elements described?''},
\emph{``Who is the intended user?''},
% \emph{``Who is meant to use it and how does it specifically cater to their needs?''},
or \emph{``What is meant by data?''}
A notable experiment in this regard is Pu \etals{}\cite{pu_probabilistic_2020} use of Algebraic Visualization\cite{kindlmann_algebraic_2014, mcnutt2021table} as a sibling to CDN, suggesting the applicability of visualization theory to DSL evaluation.

JSON DSLs have been described as being intended for use by end-users\cite{mcnutt2021integrated}. However, there has been little formal usability analysis.
Hoffswell~\etal{}\cite{hoffswell2018augmenting} studied debugging in \vega{}.
Naimipour~\etal{} \cite{naimipour2020engaging} explore social science teachers' perceived usability of \vegalite{}.
% McNutt \etal{}\cite{mcnutt2021integrated} argue that data analysts can successfully navigate the interplay between GUI and textual representation.
These works demonstrate this approach's utility. However, future work should investigate which language form is best matched with end-users.

% https://blog.marcocantu.com/blog/introducing_dsl.html
% the groups that say ````Curly-brace'' languages have limited DSL support'' are probably not referring to JSON

We believe that some of the usability issues found in JSON-style visualization DSLs\cite{liuatlas} can be addressed through careful enhancements to end-user tooling.
Merino \etal{} explore this in their system for creating notebooks tuned to individual DSLs\cite{merino2020bacat}.
Hoffswell \etal{}\cite{hoffswell2018augmenting} augment textual representations of \vega{} programs with in-situ state visualizations.
As JSON-style DSLs continue to be developed, it may be useful to explore \emph{language workbenches}\cite{fowler2010domain}, which are a form of tool for designing, composing, and using (often domain-specific) languages.
Some work has been done in this direction by JSON Schema structure editors\cite{Varv22Borowski}. However, they focus on data validation and not language design.
Some DSLs provide formal syntax definitions. However, none formalize their \emph{semantics}, although this may be because a visualization semantics language does not exist.
We suggest that a metalanguage for such descriptions would be valuable future work.

JSON-style DSLs can be \cogdim{error prone} through silent errors or overrides, such as those caused by invalid or misspelled properties. This can be confusing to the end-user who then receives little feedback on why execution is not carried out as they expect.
Tools like JSON Schema can be useful to reduce this type of error, but they are only able to capture syntactic errors. Analysis tools like linters\cite{chen2021vizlinter, hopkins_visualint_2020, mcnutt_surfacing_2020} can help capture semantic errors,
although the configuration of which may present non-trivial complexities.
Future designs should explore encoding invariants as syntax so that invalid expression is impossible.

\parahead{The Next 10k Visualization Grammars.}
If our survey prompts any prediction, it is that \emph{new visualization languages will continue to be developed}---some of which may be in the JSON-style.
There are many forms, shapes, and purposes these languages may take.
Hogr\"{a}fer \etal{}\cite{hografer2020state} argue for a map grammar.
Lau~\etal{}\cite{lau2020design} call for a computational notebook grammar that would enable task-specific notebook forms.
Hullman and Gelman call for a grammar that enables statistical model checks \cite{hullman2021designing}.
Following the trend of developing DSLs to support complex data tasks in genomics\cite{lavikkagenomespy, l2022gosling} or ML \cite{li2020p6}, other languages could be developed for other data-intensive contexts, such as multi-scale analysis, temporal data, or textual data.
Tuning computation-heavy algorithms or processes involving randomness (as in force direction) can be clumsy and error-prone, suggesting that an end-user-centered DSL enhancing those operations might be valuable.
The volume of XR DSLs suggests that JSON-style DSLs may be useful for other uncommon mediums. For instance, a sonification grammar (such as briefly explored by \highcharts{} and \dxr{}) might make non-visual data experiences easier and more accessible to produce.

Park \etal{}\cite{park2017atom} argue that efforts should be made to ``find a definitive grammar that can unify many of these existing grammars''.
However, Greenspun's tenth rule\cite{Greenspun10thProgramming} quips that any sufficiently complicated program contains an ad hoc, informally specified, bug-ridden, slow implementation of half of Lisp.
Less satirically Fowler notes that one of the biggest dangers in DSL design is ``Sliding into generality.''\cite{fowler2010domain}
While tools like \vegalite{} are probably not in danger of becoming Lisp, we suggest that consideration of small modular language components may be helpful in the continuation and extension of this ecosystem.

There are many tasks, and no one DSL will be able to capture all of them without compromising essential parts of its domain design. That is, there is \emph{no grammar to rule them all}.

\acknowledgments{
  We thank our reviewers, as well as Ravi Chugh, Arvind Satyanarayan, Brian Hempel, Will Brackenbury, and Michael A. McNutt.
}

\bibliographystyle{abbrv-doi}

\bibliography{../no-grammar}

\newcommand{\isarxiv}{X}
\ifx\isarxiv\undefined
\else
  \clearpage
  \input{./appendix}

\fi

\end{document}

%% file: commands.tex
\definecolor{linkColor}{HTML}{257E98}

\newcommand{\inlineFig}[1]{%
  \begingroup\normalfont
  \includegraphics[height=1.2\fontcharht\font`\B]{#1}%
  \endgroup
}

\usepackage{soul}
\setuldepth{Berlin}
\newcommand\asLink[2]{\textcolor{linkColor}{\href{#1}{\ul{#2}}}}

\newcommand{\liveurltext}
{https://vis-json-dsls.netlify.app/}
\newcommand{\liveurl}{\asLink{\liveurltext}{vis-json-dsls.netlify.app}}

\newcommand{\osf}{\asLink{https://osf.io/e9v8y}{osf.io/e9v8y}}

\definecolor{cogColor}{HTML}{ca2c92}
\newcommand\cogdim[1]{\textcolor{cogColor}{\emph{#1}}}

\newcommand{\etal}
{et al.}
\newcommand{\etals}
{et al.'s}
\newcommand{\eg}{{e.g.}}

\newcommand{\cf}{{cf.}}

\newcommand{\hideComments}{HIDE EM} % comment out to allow comments
\ifx\hideComments\undefined
  \newcommand\comFormat[2]{[#1: #2]}
\else
  \newcommand\comFormat[2]{}
\fi

\newcommand{\secref}[1]{\hyperref[#1]{Sec.~\ref*{#1}}}
\newcommand{\appendixref}[1]{\hyperref[#1]{Appendix.~\ref*{#1}}}
\newcommand{\figref}[1]{\hyperref[#1]{Fig.~\ref*{#1}}}
\newcommand{\eqnref}[1]{\hyperref[#1]{Eqn.~\ref*{#1}}}
\newcommand{\tabref}[1]{\hyperref[#1]{Table ~\ref*{#1}}}

{\par\kern\topsep}

\newcommand{\footnoteWithIndent}[1]
{\footnote{#1}}

\newcommand{\parahead}[1]
{%
  \vspace{0.07in}%
  \noindent%
  \textbf{\textit{#1}}%
}

\def\subsubsec#1
{\subsubsection{#1}}

\definecolor{decColor}{HTML}{795E26}
\definecolor{designColor}{HTML}{267F99}
\definecolor{deliveryColor}{HTML}{098658}

\usepackage{tikz}

\newcommand\sys[1]{{\fontfamily{phv}\selectfont{\small\textbf{#1}}}}

\newcommand\systemCite[1]{}

\newcommand{\numExamples}{4395}
\newcommand{\numLangs}{57}

\newcommand{\hideDoubles}{HIDE EM} % comment out to allow comments
\ifx\hideDoubles\undefined
  \newcommand\comboCite[2]{#1}
\else
  \newcommand\comboCite[2]{#1 #2}
\fi

%% file: system-commands.tex
\newcommand{\anivis}{\sys{AniVis}\systemCite{li2021anivis}}

\newcommand{\apexcharts}{\sys{ApexCharts}\systemCite{ApexCharts}}

\newcommand{\atom}{\sys{Atom}\systemCite{park2017atom}}
\newcommand{\bertin}{\sys{Bertin}\systemCite{bertin2022}}
\newcommand{\canis}{\sys{Canis}\systemCite{ge2020canis}}
\newcommand{\cfgConf}{\sys{CFGConf}\systemCite{devkota2021cfgconf}}
\newcommand{\chartjs}{\sys{Chart.js}\systemCite{Chartjs}}
\newcommand{\ChartML}{\sys{ChartML}\systemCite{saito2009client}}
\newcommand{\cicero}{\sys{Cicero}\systemCite{Hyeok22Cicero}}
\newcommand{\comicscript}{\sys{ComicScript}\systemCite{wang2022interactive}}
\newcommand{\compasql}{\sys{CompassQL}\systemCite{wongsuphasawat2016towards}}
\newcommand{\DataTheater}{\sys{Data Theater}\systemCite{lau2020data}}
\newcommand{\deckgljson}{\sys{deck.gl/json}\systemCite{wang2019deck}}

\newcommand{\dxr}{\sys{DXR}\systemCite{sicat2018dxr}}
\newcommand{\echarts}{\sys{ECharts}\systemCite{li2018echarts}}
\newcommand{\encodable}{\sys{Encodable}\systemCite{wongsuphasawat2020encodable}}
\newcommand{\flexer}{\sys{Flex-ER}\systemCite{lobo2020flex}}

\newcommand{\fusioncharts}{\sys{FusionCharts}\systemCite{fusioncharts}}

\newcommand{\geminitwo}{\sys{Gemini 2}\systemCite{kim2021gemini2}}
\newcommand{\genomespy}{\sys{Genome Spy}\systemCite{lavikkagenomespy}}

\newcommand{\glinda}{\sys{Glinda}\systemCite{deline2021glinda}}
\newcommand{\gosling}{\sys{Gosling}\systemCite{l2022gosling}}
\newcommand{\gotree}{\sys{GoTree}\systemCite{li2020gotree}}
\newcommand{\graphml}{\sys{GraphML}\systemCite{brandes2013graph}}
\newcommand{\gxl}{\sys{GXL}\systemCite{winter2002overview}}
\newcommand{\highcharts}{\sys{Highcharts}\systemCite{HighCharts}}
\newcommand{\ivy}{\sys{Ivy}\systemCite{mcnutt2021integrated}}
\newcommand{\jsol}{\sys{JSOL}\systemCite{yousef2022jsol}}
\newcommand{\kyrixs}{\sys{Kyrix-S}\systemCite{tao2020kyrix}}
\newcommand{\MultiClassDensityMaps}{\sys{Multiclass-Density-Maps}\systemCite{jo2018declarative}}
\newcommand{\neo}{\sys{NEO}\systemCite{gortler2021neo}}
\newcommand{\pfour}{\sys{P4}\systemCite{li2018p4}}
\newcommand{\pv}{\sys{P5}\systemCite{li2019p5}}

\newcommand{\PapARVis}{\sys{PapARVis}\systemCite{chen2020augmenting}}
\newcommand{\plotly}{\sys{Plotly JSON}\systemCite{plotlyJSON}}
\newcommand{\ScholzthreeDVisLanguage}{\sys{Scholz 3D Vis Language}\systemCite{scholz2021modular}}
\newcommand{\setcola}{\sys{SetCoLa}\systemCite{hoffswell2018setcola}}
\newcommand{\ShihVolumeVisLanguage}{\sys{Shih Volume Vis Language}\systemCite{shih2018declarative}}
\newcommand{\structgraphics}{\sys{StructGraphics}\systemCite{tsandilas_vis2021}}
\newcommand{\svl}{\sys{SVL}\systemCite{neogy2020representing}}
\newcommand{\vega}{\sys{Vega}\systemCite{satyanarayan_reactive_2016}}
\newcommand{\vegalite}{\sys{Vega-Lite}\systemCite{satyanarayan2016vega}}

\newcommand{\vizml}{\sys{VizML}\systemCite{wilkinson2012grammar}}
\newcommand{\vizql}{\sys{VizQL}\systemCite{hanrahan2006vizql}}
\newcommand{\vrviz}{\sys{VR-Viz}\systemCite{saifee2018vr}}
\newcommand{\vriajs}{\sys{VRIA}\systemCite{butcher2020vria}}
\newcommand{\xmlcharts}{\sys{XML Charts}\systemCite{XML_SWF_Charts}}
\newcommand{\zingchart}{\sys{ZingChart}\systemCite{ZingChart}}

%% file: appendix.tex
\appendix
\section{Appendix}

In this appendix we provide a complete bibliography for the languages in our survey, see \figref{fig:appendix-bib} for an index. For further information about our survey we refer the reader to the archival location of our supplement:
\osf{}
or the interactive version at \liveurl{}, which include an interactive table showing the coding, summary charts, and an interactive example explorer.

\subsection{Axes of analysis}

The following axes were used in our analysis:\\
\textbf{System}: The name of the system. \\
\textbf{Abstraction Mechanism}: Which simple abstraction mechanisms the language features, \eg{} control flow or variables \\
\textbf{Abstraction level}: Whether the language has a high (close to the domain) or low (close to the renderer) level of abstraction \\
\textbf{Allowed Data Type}: The type of input data transformed by the visualization, such as CSVs or domain-specific file formats \\
\textbf{Alt API Available}: Whether the language can be used through a mechanism other than writing programs in the language \\
\textbf{Annotation Support}: Whether there is explicit support for annotation \\
\textbf{Carrier}:  The language in which the DSL was embedded \\
\textbf{Conceptual Model}: The conceptual model which is used to represent computations executed based on the program specification \\
\textbf{Coordinate Systems}: The way in which data is arranged within the graphic \\
\textbf{Data manipulation}: The way in which language users can manipulate the input data, \eg{} filters \\
\textbf{Data model}: The conceptual model of data used by the system \\
\textbf{Dependent}: Whether the language is dependent on another for its effects \\
\textbf{Domain}:  The self described purpose of that language, such as animation, XR, genomics, etc\\
\textbf{Execution Model}: How the language is implemented, e.g. compiled, interpreted, composed \\
\textbf{Extensible}: Whether and how the language can be extended, e.g. by end users, through an API, etc \\
\textbf{Formal Definition Available}: Whether there was a formal definition of the language available (including BNF Grammars or other formal descriptions, as well as living specifications such as JSON Schemas) \\
\textbf{Interaction source}: Where the state for the application is held such that interactions with the system can be executed in a dynamic environment\\
\textbf{Juxtaposition strategy}: The way in which multiple graphics are combined, e.g. via operators \\
\textbf{Language Form}: Whether the language was internal or external \\
\textbf{Language Relationship}: The explicitly defined relationship that some languages had with one another\\
\textbf{Mark Types}: The mark types available in grammar of graphics style languages \\
\textbf{Model Formality}: Whether the model has a pervasive logical structure (high formality) or an ad hoc one (low formality) governing the syntax and conceptual model \\
\textbf{Open Source}: Whether or not the system is open source.\\
\textbf{Output Type}:  What the execution of the language produces, e.g. vector, raster, text, interactive websites\\
\textbf{Output Type}: The result of executing the program \\
\textbf{Paper}: The paper or papers describing the system. \\
\textbf{Provides Accessibility}: Whether the language explicitly provides accessibility features. \\
\textbf{Series Types}: The types of graphical series (or chart forms) available in series-based languages \\
\textbf{Source}: Where the language arises from, e.g. Academic, Industry, Open Source \\

Additional details about these axes, such as the mapping of each of the languages for each of the topics, can be found in the interactive supplement.

\input{./survey-alpha-table}

%% file: survey-alpha-table.tex
\begin{figure}[t]
  \centering
  \begin{tabular}{rp{2.8in}}
    Letter & Languages \\\hline
    A & AniVis \cite{li2021anivis}, AntVSpec \cite{antvSpec}, ApexCharts \cite{ApexCharts}, Array visualization grammar \cite{TiwariArrayVisualizationGrammar}, Atom \cite{park2017atom}\\
B & Bertin \cite{bertin2022}\\
C & Canis \cite{ge2020canis}, CFGConf \cite{devkota2021cfgconf}, Chart.js \cite{Chartjs}, ChartML \cite{saito2009client}, Cicero \cite{Hyeok22Cicero}, ComicScript \cite{wang2022interactive}, CompassQL \cite{wongsuphasawat2016towards}\\
D & Data Theater \cite{lau2020data}, deck.gl/json \cite{wang2019deck}, DGML \cite{DirectedGraphMarkupLanguage}, DotML \cite{dotML}, DXR \cite{sicat2018dxr}\\
E & ECharts \cite{li2018echarts}, Encodable \cite{wongsuphasawat2020encodable}\\
F & Flex-ER \cite{lobo2020flex}, Frappe \cite{Frappe}, FusionCharts \cite{fusioncharts}\\
G & Gemini 1 \cite{kim2020gemini}, Gemini 2 \cite{kim2021gemini2}, Genome Spy \cite{lavikkagenomespy}, gg \cite{wu2014case}, Glinda \cite{deline2021glinda}, Gosling \cite{l2022gosling}, GoTree \cite{li2020gotree}, GraphML \cite{brandes2013graph}, GXL \cite{winter2002overview}\\
H & Highcharts \cite{HighCharts}\\
I & Ivy \cite{mcnutt2021integrated}\\
J & JSOL \cite{yousef2022jsol}\\
K & Kyrix-S \cite{tao2020kyrix}\\
M & Multiclass-Density-Maps \cite{jo2018declarative}\\
N & NEO \cite{gortler2021neo}\\
P & P4 \cite{li2018p4}, P5 \cite{li2019p5}, P6 \cite{li2020p6}, PapARVis \cite{chen2020augmenting}, Plotly JSON \cite{plotlyJSON}\\
S & Scholz 3D Vis Language \cite{scholz2021modular}, SetCoLa \cite{hoffswell2018setcola}, Shih Volume Vis Language \cite{shih2018declarative}, StructGraphics \cite{tsandilas_vis2021}, SVL \cite{neogy2020representing}\\
V & Vega \cite{satyanarayan_reactive_2016}, Vega-Lite \cite{satyanarayan2016vega}, VizGrammar \cite{VizGrammar}, VizML \cite{wilkinson2012grammar}, VizQL \cite{hanrahan2006vizql}, VR-Viz \cite{saifee2018vr}, VRIA \cite{butcher2020vria}\\
X & XML Charts \cite{XML_SWF_Charts}\\
Z & ZingChart \cite{ZingChart}
  \end{tabular}
    \caption{
  Bibliographic index of languages in our survey. 
  }
  \label{fig:appendix-bib}
\end{figure}